\documentclass[preprint,12pt]{aastex}
\usepackage{amsmath}

\def\lsim {$\rlap{\raise.4ex\hbox{$<$}}\lower.55ex\hbox{$\sim$}\,$}

\newcommand{\noprint}[1]{}

\newcommand{\betabar}{\mbox{$\bar{\beta}$}}
\newcommand{\neff}{\mbox{$n_{eff}$}}
 
\newcommand{\hcop}{HCO$^+$}

\newcommand{\nthp}{N$_2$H$^+$}
 
\input{epsf}

\begin{document}

%%%%%%%%%%%%%%%%%% title %%%%%%%%%%%%%%%%%%%%%%%%%%%%%%%%%%%%%%%%
\title {\bf The Critical Density and the Effective Excitation Density of Commonly Observed Molecular
Dense Gas Tracers}
\author{Yancy L. Shirley\altaffilmark{1,2}
}
	
\altaffiltext{1}{Steward Observatory, 933 N Cherry Ave., Tucson, AZ 85721 USA}
\altaffiltext{2}{Max-Planck-Institut f\"ur Astronomie (MPIA), K\"onigstuhl 17, 69117, Heidelberg, Germany}
%\altaffiltext{3}{Adjunct Astronomer, The National Radio Astronomy Observatory}
%%%%%%%%%%%%%%%%%% abstract %%%%%%%%%%%%%%%%%%%%%%%%%%%%%%%%%%%%%%%%
 
\begin{abstract}
The optically thin critical densities and the effective excitation densities to 
produce a 1 K km/s (or $0.818$ Jy km/s $(\frac{\nu_{jk}}{100 \rm{GHz}})^2 \, (\frac{\theta_{beam}}{10^{\prime\prime}})^2$) spectral line 
are tabulated for 12 commonly observed dense gas molecular tracers.  
The dependence of the critical
density and effective excitation density on physical assumptions (i.e. gas kinetic temperature 
and molecular column density) is analyzed.  Critical densities 
for commonly observed dense gas transitions in molecular clouds 
(i.e. HCN $1-0$, HCO$^+$ $1-0$, N$_2$H$^+$ $1-0$)
are typically $1 - 2$ orders of
magnitude larger than effective excitation densities because 
the standard definitions of critical density do not account for radiative trapping
and $1$ K km/s lines are typically produced 
when radiative rates out of the upper energy level of the transition 
are faster than collisional depopulation. 
The use of effective excitation density has a distinct advantage over the use
of critical density 
in characterizing the  differences in density traced by species such as 
NH$_3$, HCO$^+$, N$_2$H$^+$, and HCN as well as their isotpologues; 
but, the effective excitation density
has the disadvantage that it is undefined for transitions when $E_u/k \gg T_k$, for
low molecular column densities, and for heavy molecules with complex 
spectra (i.e. CH$_3$CHO).
\end{abstract}

%\keywords{ISM: dense molecular gas}

%%%%%%%%%%%%%%%%%% 1. Main Text %%%%%%%%%%%%%%%%%%%%%%%%%%%%%%%%%%%%%%%%

\section{Introduction}

The density at which a particular molecular transition is excited has always been
an interesting quantity for studies of molecular gas in the interstellar medium.
Two concepts have been developed, the critical density and the effective excitation
density, to quantify the excitation density of a transition.  
A traditional approach in studying the critical density has been to consider
the molecular rotational energy levels as a simple two level system.  While
this approach has pedagogical advantages, the interpretation of
concepts, such as critical density, are oversimplified and incorrect.  Molecules
are fundamentally multi-level systems with radiative and collisional processes between 
the levels (Figure 1).  It is almost never appropriate to approximate commonly observed
dense gas tracers, such as \hcop , HCN, NH$_3$, \nthp , etc., as two level systems.
Therefore, in this tutorial, we eschew the two level approximation and consider
the full multi-level nature of molecular excitation.  It is assumed
that the reader has a basic familiarity with statistical equilibrium, radiative transfer,
and molecular spectroscopy.  For a summary of radiative transfer techniques and
molecular data used in those calculations, see van der Tak (2011) and references therein.
The purpose of this tutorial is to quantify and 
systematize the use of critical density and effective excitation density for
dense gas molecular tracers.

We shall use the excitation of two molecules, \hcop\ and NH$_3$, as examples throughout
this tutorial.  Figure 1 shows the energy level structure for both molecules.  
\hcop\ is an abundant linear molecule with a simple $^{1}\Sigma$ electronic ground state
meaning that there is zero net electronic spin and zero net electronic angular momentum that
could interact with and split the rotational levels of the molecule.  Electric dipole
selection rules limit transitions to $\Delta J = \pm 1$.
The $1-0$ ground state rotational transition is a millimeter transition at 89 GHz (3.3 mm)
and the frequency of rotational transitions progress linearly into the submillimeter
part of the spectrum.

NH$_3$ is an abundant pyramidal 
symmetric top molecule with a simple $^{1}A_1$ electronic ground state also
meaning that there is zero net electronic spin and zero net electronic angular momentum.
The energy levels are designed by $J_K$ where $J$ is the total rotational angular
momentum and $K$ is the projection of the angular momentum 
onto the NH$_3$ symmetry axis.  Electric dipole
transition do not allow the projected angular momentum to change ($\Delta K = 0$).
As a results, the energy level structure is usually organized in K-ladders with each
K-ladder designated as either ortho ($K$ = 0, 3, 6, etc.) or para ($K$ = 1, 2, 4, 5, etc.)
due to the spin symmetry required to satisfy Fermi-Dirac statistics 
for each rotational energy level.  The plane of three
H atoms can tunnel from positions above and below the N atom resulting in inversion splitting
of each rotational level (with symmetry denoted by $\pm$) 
except for levels with $K = 0$ for which half of the inversion levels
are missing due to the Pauli exclusion principle (see Townes \& Schwalow 1975 for
a detailed explanation).  Electric dipole transitions are allowed between the inversion-split levels
with different symmetries and these inversion transitions occur at wavelengths near 1.3 cm
(24 GHz). 
Since $J$ and $K$ do not change in a pure inversion transition,
the shorthand notation $(J,K)$ has been adopted.
As with \hcop , there are rotational transition with $\Delta J = \pm 1$, but since NH$_3$ is a 
lighter molecule than \hcop\ ($E_{rot} \propto 1/I$ where I is the moment of inertia of
the molecule measured from a coordinate system at the center-of-mass), 
the transitions occur at submillimeter and far-infrared
wavelengths.  Thus, the lowest energy transitions of \hcop\ and NH$_3$ cover a wide range of
wavelengths from the centimeter to the far-infrared.

%%%%%%%%%%%%%%%%%% Figure 1  %%%%%%%%%%%%%%%%%%%%%%%%%

\begin{figure}
\figurenum{1}
\epsscale{1.0}
\plottwo{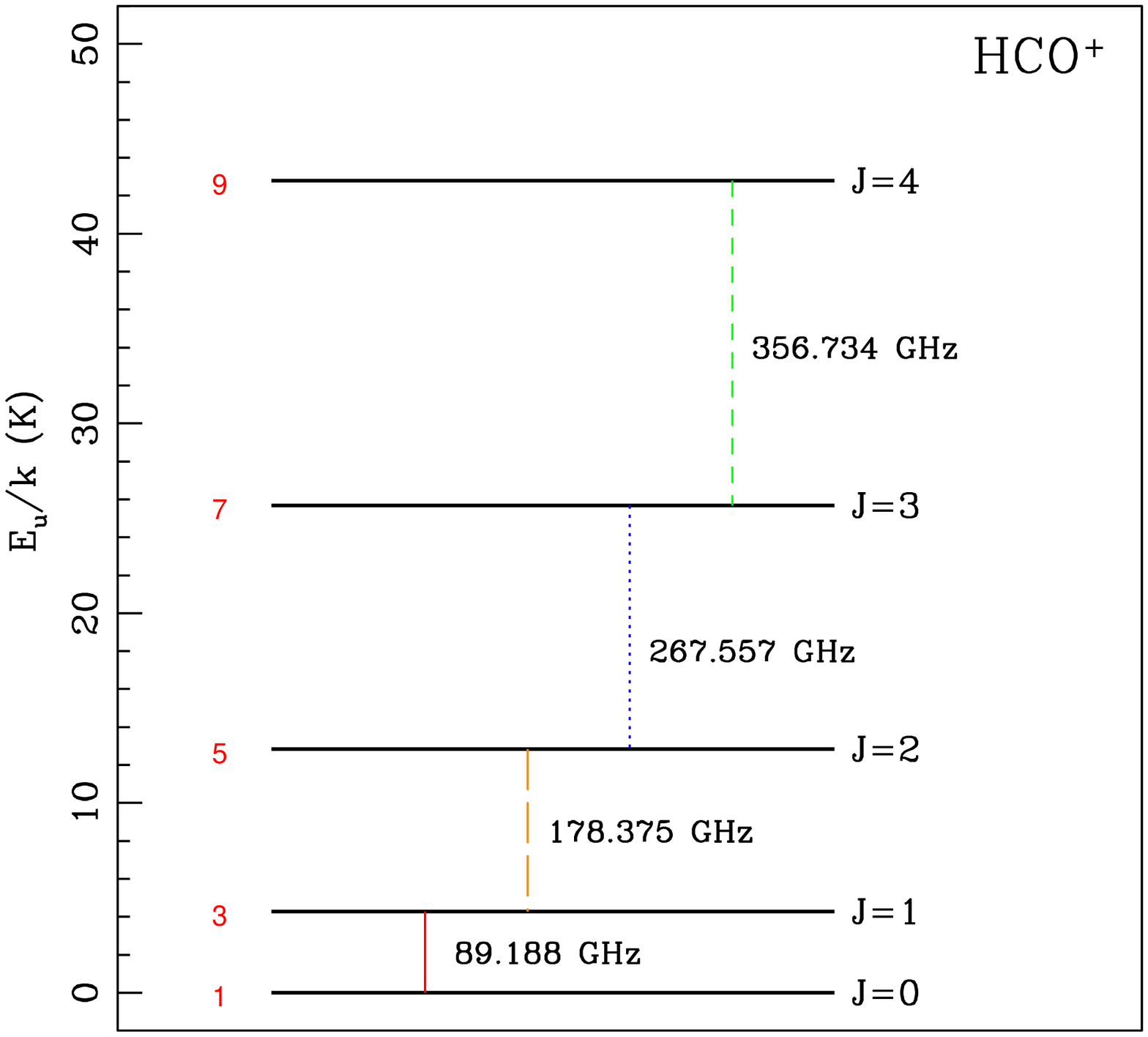}{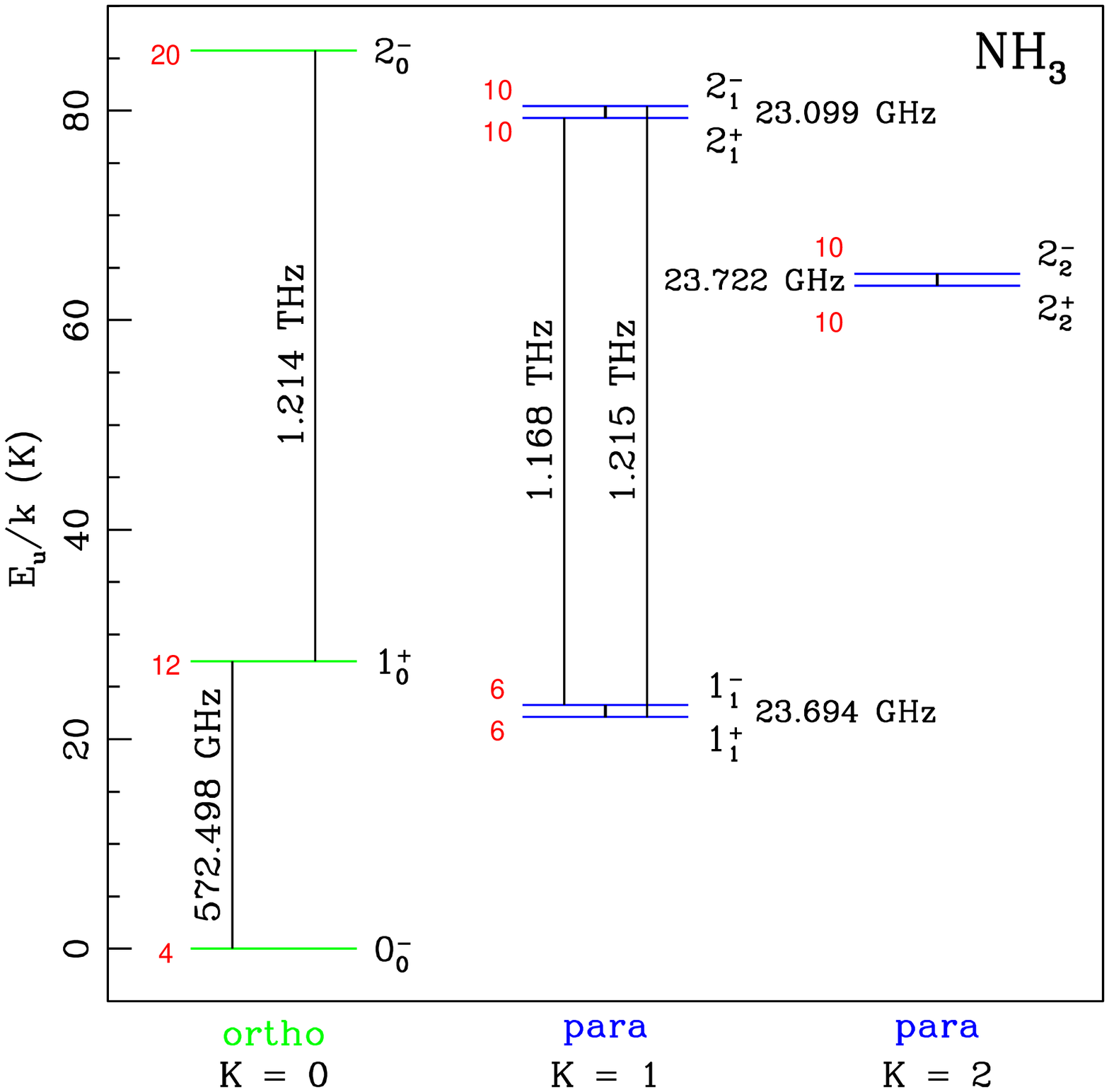}
\figcaption{
Left: The energy level structure for \hcop\ showing the lowest $4$ energy levels.
Electric dipole allowed transitions are shown along with the quantum numbers
($J$) to the right of each level and statistical weights in red to the left of each energy level.
Right:  The energy level structure for NH$_3$ showing the lowest $9$ energy levels.
Electric dipole allowed transitions
are shown along with the quantum numbers ($J_K^{\pm}$) to the right of each level 
and statistical weights in red to the left of each energy level.  This figure was
adapted from figures in Mangum \& Shirley (2015).}
\end{figure}

\section{Critical Density}

The critical density has traditionally been used as a measure of the density
at which a particular transition is excited and is observed at radio wavelengths. The definition
of the critical density is not consistent throughout the
literature.  Some definitions only consider the two energy levels involved in the transition
(2 level approximation) while other definitions use the multi-level nature of collisions
to sum over all collisions out of the upper energy level or only from the upper
energy level to lower energy levels.  All of the standard definitions assume the molecular
emission is optically thin and ignore radiative trapping which 
is very important and cannot be ignored for many commonly observed dense gas tracers
(e.g. HCN $1-0$, HCO$^+$ $1-0$, N$_2$H$^+$ $1-0$, etc.; see \S2.2).  The standard
definitions also usually ignore continuum backgrounds (see \S4).

We start by defining the critical density as the density
for which the net radiative decay rate from $j \rightarrow k$
equals the rate of
collisional depopulation out of the upper level $j$ for a multilevel system.  
The net radiative decay rate from $j \rightarrow k$ is the
spontaneous decay rate from $j \rightarrow k$ plus the stimulated emission rate
from $j \rightarrow k$ minus the absorption rate from $k \rightarrow j$. 
Mathematically, the ratio of rates is
\begin{equation}
\Re = \frac{n_j A_{jk} + n_j B_{jk} u_{\nu_{jk}} - n_k B_{kj} u_{\nu_{jk}}}
{n_{c} n_j \sum_{i \neq j} \gamma_{ji}} \;\; ,
\end{equation}
where
$n_j$ is the number density (cm$^{-3}$) of molecules in the upper level of the transition,
$n_k$ is the number density (cm$^{-3}$) of molecules in the lower level of the transition,
$n_c$ is the number density (cm$^{-3}$)  of colliding particles (typically H$_2$, H, or e$^-$),
and $\gamma_{ji}$ are the collision rates (cm$^3$ s$^{-1}$) out of level $j$ into 
another level $i$.
$n_c$ is equal to $n_{crit}$ when $\Re = 1$.
$A_{jk}$ and $B_{jk}$ are the Einstein $A$ (s$^{-1}$) and $B$ 
(erg$^{-1}$ cm$^{3}$ s$^{-2}$) coefficients related by 
\begin{equation}
\frac{g_k}{g_j} B_{kj} = B_{jk} = \frac{c^3}{8\pi h\nu^3_{jk}} A_{jk} \;\;\; .
\end{equation} 
$u_{\nu_{jk}}$ is the energy density (erg cm$^{-3}$ Hz$^{-1}$) of the radiation fields
at the frequency of the transition $\nu_{jk}$.  The energy density is 
written in terms of a Planck function and is related to the angle and
polarization averaged photon occupation number, $n_{ph}$, by
\begin{equation}
u_{\nu}(T) = \frac{8\pi h \nu^3}{c^3} \frac{1}{e^{h\nu/kT} - 1} = \frac{8\pi h \nu^3}{c^3} n_{ph}(T,\nu) \;\;\;.
\end{equation} 
The total energy density can include contributions from
the cosmic microwave background (CMB), dust continuum emission from 
the interstellar radiation field, dust continuum emission local to the molecular cloud, 
optically thick molecular emission at the frequency $\nu_{jk}$, 
and other continuum and spectral line sources.

\subsection{Optically Thin Approximation}

We define the optically thin critical density without a background by solving for $n_c$
in Equation 1 when $\Re = 1$ and ignoring the stimulated emission and absorption terms:
\begin{equation}
n_{crit}^{thin, no\; bg} = \frac{ A_{jk} }{\sum_{i \neq j} 
\gamma_{ji}} = 
\frac{ A_{jk}}{\sum_{i < j} \gamma_{ji} + \sum_{i > j}
\frac{g_i}{g_j} \gamma_{ij} e^{-(E_i - E_j)/kT_k} } \;\;\; .
\end{equation}
The sum over collision rates in the denominator of Equation 4
has been split in two terms depending on whether the collision from level $j$ is to an
energy level lower in energy (first sum) or higher in energy (second sum) than level $j$.  
Upward collision rates are related to downward collision rates using detailed
balance by 
\begin{equation}
\gamma_{ji} = \gamma_{ij} \frac{g_i}{g_j} e^{-(E_i - E_j)/kT_k} \;\; \rm{for} \;\; E_i > E_j
\end{equation}
where $T_k$ is the kinetic temperature of the collisional partner.
Published collision rates (i.e. the Leiden molecular database\footnote{http://home.strw.leidenuniv.nl/$\sim$moldata/ }; Sch{\"o}ier et al. 2005) 
quote only downward collision rates. 
In the two level approximation, which is widely quoted but rarely appropriate, 
the sums reduce to a single term such that 
$n_{crit, 2lvl}^{thin,no\; bg} = \frac{A_{jk}}{\gamma_{jk}}$.  As a result,
$n_{crit, 2lvl}^{thin,no\; bg}$ will always be larger (usually by at least a factor of a few)
than $n_{crit}^{thin, no\; bg}$.

%%%%%%%%%%%%%%%%%% Figure 2  %%%%%%%%%%%%%%%%%%%%%%%%%

\begin{figure}
\figurenum{2}
\epsscale{1.0}
\plotone{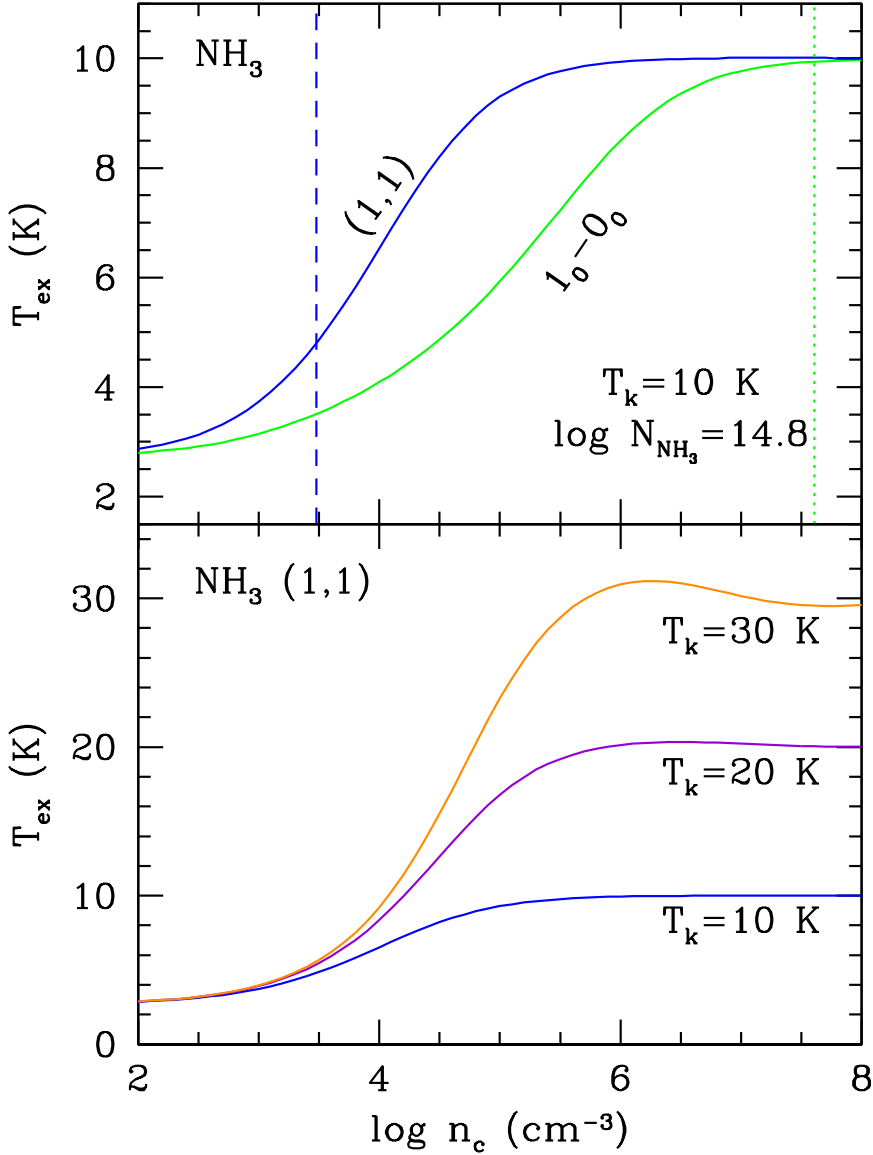}
\figcaption{
Top: The excitation temperature is plotted for the NH$_3$ (1,1) inversion
transition (23.7 GHz, 1.3 cm) 
and $1_0-0_0$ rotational transition (572.5 GHz, 0.52 mm) for a gas kinetic 
temperature of $10$K and a total NH$_3$ reference column density of 
$\log N = 14.8$ cm$^{-3}$ (an ortho to para ratio of $2:1$ is assumed).  
The blue dashed line and the green dotted line indicate $n_{crit}^{thin, no\; bg}$
for the (1,1) and $1_0-0_0$ transitions respectively.
Bottom: The variation of excitation temperature for the NH$_3$ (1,1) transition
with gas kinetic temperature is plotted for $T_k =$ 10, 20, and 30 K.}
\end{figure}

It is instructive to compare how the optically thin critical density compares to the excitation temperature.
For a particular transition between levels $j \rightarrow k$, 
the level population are related to the excitation temperature, $T_{ex}$, between 
the two levels by Boltzmann's equation
\begin{equation}
\frac{n_j}{n_k} = \frac{g_j}{g_k} \exp\left( - \frac{E_j - E_k}{kT_{ex}} \right) \;\; \rm{for} \;\; E_j > E_k \;\; .
\end{equation}
The excitation temperature (and therefore the level populations) depend on the
density of colliding partners ($n_c$), the gas
kinetic temperature of colliding partners ($T_k$), and the energy density of radiation
fields that the molecules are exposed to.  The dependence of $T_{ex}$
on the density and gas kinetic temperature for the ground state
transitions of ortho and para NH$_3$ are
shown in Figure 2.  The only background radiation field assumed in this
example is the CMB.  At low densities ($n_c < 10^3$ cm$^{-3}$), the level populations
come into equilibrium with the CMB and $T_{ex}$ approaches $T_{cmb} = 2.725$ K.
As the density increases, collisions become more important in determining
the level populations.  The excitation temperature of $(1,1)$ centimeter wavelength 
inversion transition equilibrates to $T_k$ by densities $> 10^{5.5}$ cm$^{-3}$
while the submillimeter $1_0 - 0_0$ rotational transition equilibrates to
$T_k$ at significantly higher densities ($> 10^7$ cm$^{-3}$).  The general shape
of the $T_{ex}$ with $\log n_c$ curves is a sigmoid curve with stimulated
coupling with the background radiation field setting the lower bound at low 
densities and collisions with colliding partners at $T_k$ setting the upper bound.
At densities where $T_{ex} < T_k$, the level populations are said to be
sub-thermally populated.  This is the situation for most of the commonly observed
dense gas transitions in the ISM.  
When $T_{ex}$ approaches $T_k$, the level populations are said to be thermalized.\footnote{Note that the $T_{ex}$ curve for NH$_3$ (1,1) shows a region of super-thermal
excitation ($T_{ex} > T_k$) at densities immediately before thermalization for
the $T_k = 30$ K curve. 
This is due to a bottleneck in radiative transitions, caused by the five orders of
magnitude difference between the spontaneous emission rates for
the inversion transition and the THz rotational transition into the upper
inversion level, resulting in a slight overpopulation of the upper level of
the inversion transition.  This effect is more pronounced for higher $T_k$
because upward collisions to the $J_K = 2_1$ levels are more efficient at high $T_k$
(see Equation 5).}

At optical wavelengths (the high frequency limit), critical density 
is interpreted as the density at which an atomic forbidden line is quenched by 
collisions meaning that a collisional de-excitation occurs before a photon can be generated from
spontaneous emission.  In the optical, quenching occurs when $T_{ex}$ is thermalized
($T_{ex} \rightarrow T_{k}$).
At radio wavelengths (the low frequency limit) 
the critical density
occurs at different positions along the sigmoid $T_{ex}$ curve for different molecular tracers
and different transitions.
In the case of the 1.3 cm (23.7 GHz) 
NH$_3$ (1,1) transition, the optically thin critical density 
is two orders of magnitude below the thermalization
density (Figure 2).  Early radio astronomy observations of molecules such as OH and NH$_3$
occurred at centimeter wavelengths and since the critical density at these low
frequencies is just above the regime where the $T_{ex}$ starts to rise
above equilibration with the CMB, critical density was used to indicate the density
at which a transition would appear in emission.  As the frequency of the transition
increases, the critical density moves farther up the $T_{ex}$ curve.
For the $0.52$ mm ($572$ GHz) NH$_3$ $1_0 - 0_0$ transition, the critical density is nearly
equal to the density required for thermalization.  Millimeter transitions are an
intermediate case; thus, neither the radio interpretation of critical density being
the density at which an emission line is excited nor the optical interpretation
of critical density being the density at which an emission line is quenched is appropriate.

\subsection{Optically Thick Emission}

In reality, radiative trapping is very important for many molecular transitions observed 
in the dense interstellar medium and cannot be ignored.   The energy
density is then modified using the solution to the equation of radiative transfer 
to include emission from the molecule at the excitation temperature describing the 
level populations for a transition from $j \rightarrow k$ as
\begin{equation} 
u_{\nu_{jk}} = u_{\nu_{jk}}(T_{bg})\bar{\beta} + u_{\nu_{jk}}(T_{ex})(1 - \bar{\beta}) \;\;\; ,
\end{equation}
where \betabar\ is the solid angle averaged escape fraction
(see Chapter 19 of Draine 2011 for a derivation). If we multiply the energy density 
by the conversion factor between Einstein $B_{jk}$ and $A_{jk}$
terms from Equation 2, then we find that 
\begin{equation}
\frac{c^3}{8\pi h\nu^3_{jk}} u_{\nu_{jk}} = n_{ph}(T_{bg},\nu_{jk})\bar{\beta} 
+ n_{ph}(T_{ex},\nu_{jk})(1 - \bar{\beta}) \;\; .
\end{equation}
In the limit that 
the background radiation can be ignored ($u_{\nu_{jk}}(T_{bg})\bar{\beta} \ll  u_{\nu_{jk}}(T_{ex})(1 - \bar{\beta})$), 
the critical density becomes
\begin{eqnarray}
n_{crit}^{thick, no\; bg}  & = &  \frac{ n_j A_{jk} + n_j A_{jk} n_{ph}(T_{ex},\nu_{jk}) (1 - \bar{\beta}) 
- n_k \frac{g_j}{g_k} A_{jk} n_{ph}(T_{ex},\nu_{jk})(1 - \bar{\beta)}}
{n_j \sum_{i \neq j} \gamma_{ji} }
\nonumber \\
 & = &
\frac{ n_j A_{jk} \left[ 1 + (1 - \frac{g_j}{g_k}\frac{n_k}{n_j})n_{ph}(T_{ex},\nu_{jk})(1 - \bar{\beta}) \right] }
{n_j \sum_{i \neq j} \gamma_{ji}  }   \nonumber \\
& = & 
\frac{ A_{jk} \left[ 1 + (\frac{-1}{n_{ph}(T_{ex},\nu_{jk})})n_{ph}(T_{ex},\nu_{jk})(1 - \bar{\beta}) \right] }
{\sum_{i \neq j} \gamma_{ji}  }   \nonumber \\
& = & 
\frac{ \bar{\beta} A_{jk} }
{ \sum_{i \neq j} \gamma_{ji} } \;\;.
\end{eqnarray}
Comparing Equation 9 with Equation 4, we see that 
the effect of radiative trapping is to reduce the spontaneous transition 
rate by the escape 
fraction \betabar $A_{jk}$ (see Scoville \& Solomon 1974, Goldreich \& Kwan 1974).

The relationship between \betabar\ and the line of sight optical depth, $\tau_{\nu_{jk}}$, depends on the geometry and kinematics
of the region. 
In the case of a static, uniform density sphere
\begin{equation}
\bar{\beta} = \frac{3}{4\tau_{\nu_{jk}}} - \frac{3}{8\tau^3_{\nu_{jk}}} + e^{-2\tau_{\nu_{jk}}}\left( \frac{3}{4\tau^2_{\nu_{jk}}} 
+ \frac{3}{8\tau^3_{\nu_{jk}}} \right) 
\end{equation} 
(Osterbrock 1989), while 
in the case of a spherical cloud with a large velocity gradient $v \propto r$ 
(the LVG or Sobolev approximation),
\begin{equation}
\bar{\beta} = \frac{(1 - e^{-\tau_{\nu_{jk}}})}{\tau_{\nu_{jk}}} 
\end{equation} 
(Sobolev 1960, Castor 1970).  
For large optical depths ($\tau_{\nu_{jk}} >\sim 5$), 
the effective spontaneous transition rate is then given by 
$\bar{\beta}A_{jk} \sim A_{jk}/\tau_{\nu_{jk}}$.

This result has an important effect on the critical density by reducing its
value from the optically thin value.  The most commonly observed transition
of molecular gas is $^{12}$CO $1-0$.  The optically thin 
critical density at $T_k = 10$ K is $n_{crit}^{thin, no\; bg} = 1 \times 10^3$ cm$^{-3}$;
however, since $^{12}$CO $1-0$ has optical depths that are typically 
more than a factor of $10$ in molecular clouds (and usually much higher), 
it is easy to observe 
strong lines ($> 1$ K km/s) in gas that has densities $< 10^2$ cm$^{-3}$.
Optical trapping is so important for CO that it does not make
sense to use the optically thin critical density.

Commonly observed lines of dense gas tracers in cores and clumps also 
tend to be optically thick.
The upper left panel of Figure 3 shows the predicted
optical depth of \hcop\ $1-0$ through $4-3$ 
at different densities from a simple (single density,
single temperature) radiative transfer model predicting that these
transition are expected to be very optically thick over a wide
range of densities.  Observations support this prediction.
For example, several hundred high-mass clumps mapped in the MALT90 Galactic survey 
have a median optical depth of $\tau = 23$ for the \hcop\ $1-0$ transition
(Hoq et al. 2013).  Clumps observed in the Bolocam Galactic Plane Survey (BGPS) 
in the higher frequency transition \hcop\ $3-2$ have an average $\tau = 10$
(Shirley et al. 2013).  
As a result, the analytically calculated 
critical densities using the optically thin approximation (Equation 4)
are systematically biased toward higher densities by an order of magnitude 
or more.  \hcop\ is not unique as most commonly observed dense gas tracers
(i.e. HCN, CS, N$_2$H$^+$, etc.)
in the dense, cold neutral medium are typically both sub-thermally populated 
($T_{ex} < T_k$) and optically thick (also see Plume et al. 1997, Gerner et al. 2014). 
In practice, to properly estimate
the optical depth, a coupled radiative transfer and
statistical equilibrium calculation must be performed to find the
level populations upon which the optical depth depends
(since $\tau_{\nu_{jk}} \propto \int (n_k B_{kj} - n_j B_{jk}) \, ds$ along 
the line-of-sight $s$). 
Since the escape fraction depends nonlinearly 
with column density (for modest optical depths), nonlinearly with geometry, and 
nonlinearly with gas kinetic temperature,
it is difficult to make a simple
tabulation of a \betabar\ correction to $n_{crit}^{thin, no\; bg}$.

%%%%%%%%%%%%%%%%%% Figure 3  %%%%%%%%%%%%%%%%%%%%%%%%%

\begin{figure}
\figurenum{3}
\epsscale{1.0}
\plotone{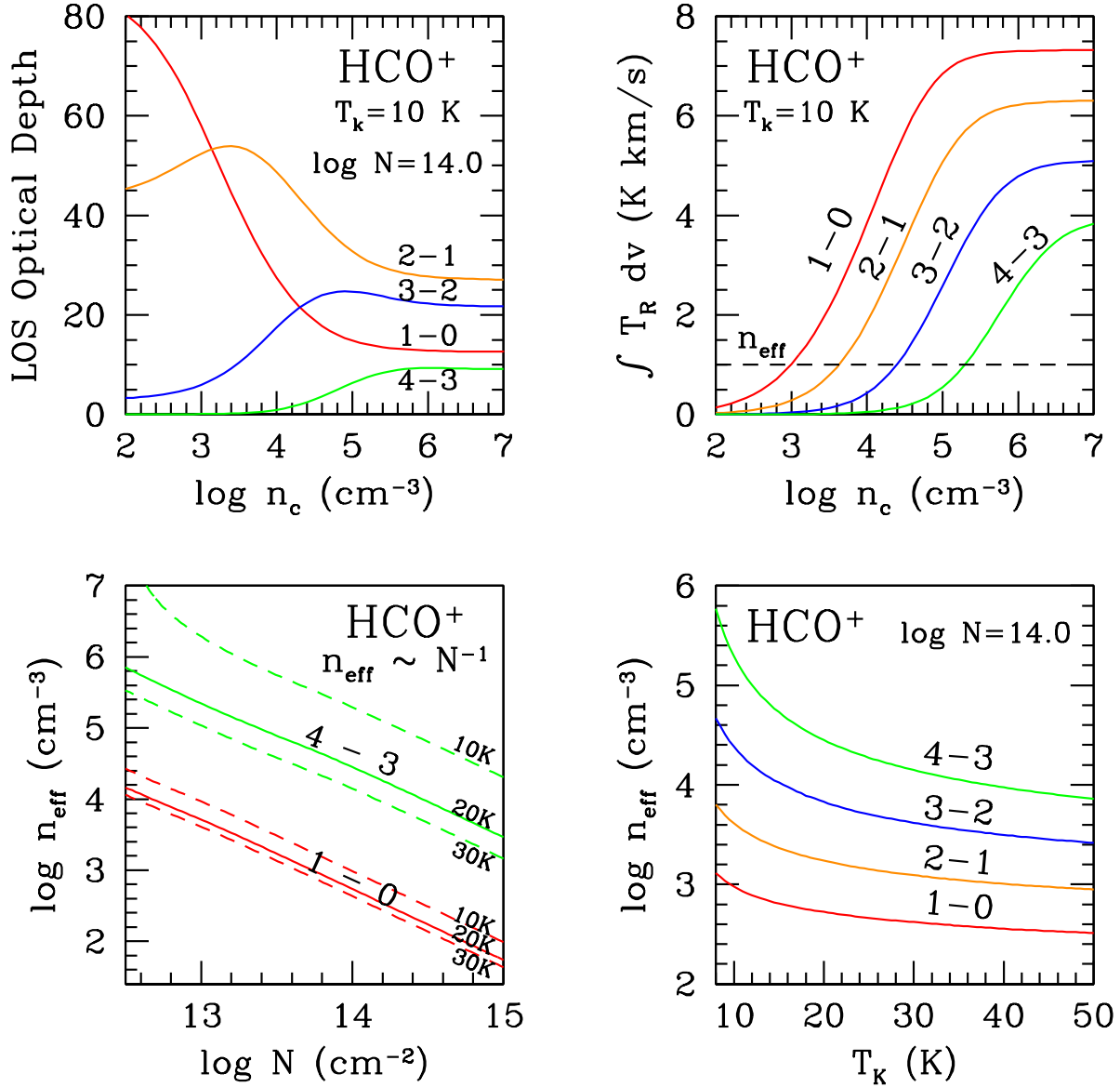}
\figcaption{ 
Top Left: Line of sight optical depth for \hcop\ transitions.
A column density of $\log N = 14.0$ and a gas kinetic temperature
of $T_k = 10$ K were assumed.
Top Right: The integrated intensity for \hcop\ transitions.
A column density of $\log N = 14.0$ and a gas kinetic temperature
of $T_k = 10$ K were assumed.
The horizontal dashed line is the criterion for the definition of
\neff\ used in this paper.
Bottom Left: $n_{eff}$
is inversely proportional column density.  
Only the \hcop\ $1-0$ and $4-3$ transitions
are shown for clarity.  The solid line is calculated at $T_k = 20$ K while the dashed
lines above and below the solid line correspond to $T_k = 10$ K and $T_k = 30$ K. 
Bottom Right: The dependence of
$n_{eff}$ on choice of gas kinetic temperature for the four lowest rotational transitions
of \hcop .  $n_{eff}$ was calculated assuming the reference column density of 
$\log N = 14.0$.}
\end{figure}

\section{Effective Excitation Density}

It is possible to detect strong ($1$ K km/s) molecular lines at densities
well below the critical density (see Figure 3). 
An alternative definition has been developed to
trace the effective density at which a modest line intensity is observed that is
based on radiative transfer calculations with reasonable assumptions 
about the column density and gas kinetic temperature of the region (Evans 1999).
The effective excitation density is defined as
\begin{equation}
n_{eff} = n_c \; \ni \; \int T_R \, \mathrm{dv} = 1 \; \rm{K} \, \rm{km/s} \;\;\; ,
\end{equation}
the density which results in a molecular line with an integrated intensity of $1$ K km/s.
The choice of $1$ K km/s is arbitrary but represents an easily detectable integrated intensity
\footnote{The radiation temperature is related to the observed flux density $S_{\nu_{jk}}$ 
through $T_R = \frac{c^2}{2k\nu_{jk}^2\Omega_{beam}} \, S_{\nu_{jk}}$ where 
$\Omega_{beam} = \frac{\pi \theta_{beam}^2}{4 \ln 2}$ is the solid angle of the telescope
beam with FWHM $\theta_{beam}$.  A $1$ K km/s integrated intensity corresponds to
an integrated flux density of $0.818$ Jy km/s $(\frac{\nu_{jk}}{100 \rm{GHz}})^2 \, (\frac{\theta_{beam}}{10^{\prime\prime}})^2$.}.
For lower integrated intensity criteria, \neff\ decreases (Figure 3).  
It is important to note
that the effective excitation density is based on a simple empirical criterion and is not
based on $\Re = 1$ or any specific ratio of radiative rates 
to collisional depopulation rates.  \neff\ does however account for the effects of
radiative trapping.
The original definition presented in Evans (1999) defined $n_{eff}$ as the density
resulting in $T_R = 1$ K line and used a LVG radiative transfer code to calculate
$n_{eff}$.  
We shall use the radiative transfer code RADEX 
(van der Tak et al. 2007) 
with the escape probability formalism for a static, uniform density sphere,
with the CMB as the only background radiation field, and with H$_2$ as the
dominant collision partner for all calculations.
The definition in Equation 12 is approximately equivalent to the Evans 1999 definition 
when a FWHM linewidth of 1 km/s is assumed in RADEX calculations.

\subsection{Properties of Effective Excitation Density}

The column density of the observed molecule
and the gas kinetic 
temperature of the colliding partner must be assumed in order to calculate \neff .  
The sensitivity of \neff\ to these assumptions is explored in Figure 3 for the 
\hcop\ molecule.
At lower column densities, a higher \neff\ is required to produce a 
$1$ K km/s line.  As the column density increases by an order of magnitude, the \neff\ 
decreases by almost exactly an order of magnitude.  
The effective excitation density is inversely proportional to column density 
\begin{equation}
n_{eff}(T_k,N) \approx n_{eff}(T_k,N_{ref}) \, \frac{N_{ref}}{N} \;\;\; ,
\end{equation}
where $N_{ref}$ is a reference column density.  This simple scaling
proportionality breaks down for $n_{eff} \gtrsim 5 \times 10^6$ cm$^{-3}$
and the relationship between \neff\ and $N$ becomes noticeably non-linear
(see the $T_k = 10$ K curve for \hcop\ $4-3$ in the bottom left panel of Figure 3).  
In those cases, a new RADEX model should be run to scale tabulated results
to a new column density.

The dependence of \neff\ to $T_k$ is nonlinear and 
more sensitive at low kinetic temperatures ($T_k < 20$ K)
due to the inefficiency of upward collisions at lower $T_k$ (Figure 3).  
An order of magnitude higher 
\neff\ is required to make a $1$ K km/s line for $T_k = 10$ K than
at $T_k = 30$ K for \hcop\ $3-2$.  This sensitivity is due to the exponential dependence
from upward collisional rates (Equation 4) and becomes more pronounced when the upper energy level
is significantly 
higher than $T_k$ (i.e. the effect is stronger for $4-3$ vs. $1-0$ at $T_k < 20$ K 
because $E_u/k = 42.8$ K for J = 4 vs. $E_u/k = 4.3$ K for J = 1).

Table 1 gives $n_{eff}$ for $12$ dense gas tracers 
at $T_k$ = 10, 15, 20, 50, and 100 K for a reference
column density.
Recent surveys of dense gas in the Milky Way may be used to guide
our choice of the reference column density.  For example, observations of several hundred clumps
by the MALT90 (Meittenen et al. 2013, Hoq et al. 2013), ChaMP (Barnes et al. 2012), 
and BGPS (Shirley et al. 2013) surveys indicate that \hcop\ column densities 
typically span from $\log N = 12 - 15.5$.  The Hoq et al. and Shirley et al. surveys
find median column densities of clumps drawn from the ATLASGAL and BPGS
Galactic plane surveys that are in agreement near $\log (\rm{med}\{N\}) = 14.4$.  
Miettenen's (2014) analysis of infrared dark clouds (IRDCs) indicates a significantly
lower median column density of $\log (\rm{med}\{N\}) = 13.1$.  All of these
surveys use observations of H$^{13}$CO$^+$ to correct the optical depth of the \hcop\ line.
The Barnes et al. (2011) blind mapping survey of a section of the southern Galactic plane find a 
smaller value of $\log (\rm{med}\{N\}) = 12.9$, but did not observe H$^{13}$CO$^+$
to make optical depth corrections.  
%All of these surveys have focused on massive clumps.
We chose a column density that is intermediate
between the IRDC, low-mass samples, and the Galactic plane surveys of 
$\log (N_{ref}) = 14.0$.  For comparison, Evans (1999) tabulate the effective density for
\hcop\ assuming $\log N = 13.5$.

The reference column densities for each molecule in Table 1 are estimated 
from surveys in dense gas tracers  (Suzuki et al. 1992, Zylka et al. 1992,
Mangum \& Wootten 1993, Plume et al. 1997, Jijina et al. 1999, Savage et al. 2002,
Pirogov et al. 2003, Purcell et al. 2006, Blair et al. 2008, Rosolowsky et al. 2008,
Falgarone et al. 2008, Hily-Blant et al. 2010, Barnes et al. 2011, Dunham et al. 2011, 
Ginsburg et al. 2011, Reiter et al. 2011,
Adande et al. 2012, Benedettini et al. 2012, Wienen et al. 2012, 
Hoq et al. 2013, Shirley et al. 2013, Gerner et al. 2014,  Miettenen et al. 2014
Yuan et al. 2014).  Since these surveys contain many biases (i.e. only
study massive high-mass regions or only study
nearby low-mass regions, etc.), $N_{ref}$ represents an educated guess at a typical value 
intermediate between high-mass clumps and low-mass cores. 
The tabulated $n_{eff}$ can be easily scaled to another column density using 
Equation 13.   
$^{13}$C isotopologue column densities assume
no fractionation and an isotope ratio of $[^{12}\rm{C}]/[^{13}\rm{C}] = 50$,
appropriate for a source in the Molecular Ring (Wilson \& Rood 1994).
For molecules with ortho and para states, LTE statistical equilibrium was
assumed ($2:1$ ortho to para ratio for NH$_3$ and $3:1$ ortho to para ratio for
H$_2$CO).

\subsection{Comparing Critical Density and Effective Excitation Density}

The panels of 
Figure 4 summarize the range in $n_{crit}^{thin, no\; bg}$ and $n_{eff}$ of commonly observed 
molecular tracers of dense gas for reference observed column densities and a gas kinetic temperature
of $T_k = 15$ K.  This temperature was chosen to be intermediate between the
typical temperature of nearby low-mass cores ($10$ K, Jijina et al. 1999, Rosolowsky et al.
2008, Seo et al. 2015, submitted) 
and high-mass clumps in the Galactic plane ($20$ K, Dunham et al. 2011, Wienen et al. 2012).
$n_{eff}$ is lower than $n_{crit}^{thin, no\; bg}$ in nearly all cases by $1 - 2$ orders of
magnitude.  This is partially due to the sensitivity of $n_{eff}$ to line trapping in optically
thick transitions.  Transitions that are the most optically thick 
have the largest average $n_{crit}^{thin, no\; bg} / n_{eff} $ ratios (i.e
$121$ for \hcop\ transitions and $88$ for HCN transitions where the
average is calculated for $1-0$ through $4-3$ transitions).  
Less optically thick species have smaller ratios
(i.e. average ratios of $16$ for CN and $10$ for N$_2$H$^+$).
The species plotted with the smallest difference between 
$n_{crit}^{thin, no\; bg}$ and $n_{eff}$ is the optically thin isotpologue H$^{13}$CN
with an average ratio of $1.7$.  
%$n_{eff}$ is less than $n_{crit}^{thin, no\; bg}$
%for all H$^{13}$CN transitions with the single exception of the $4-3$ transition. 
Some heavier species whose transitions are likely close to
optically thin listed in Table 1 also have low average ratios that are less than $3$ (i.e. 
H$_2$CO, HC$_3$N, and CH$_3$CN).
These results indicate that a $1$ K km/s line is typically observed for transitions when
$\Re > 1$ in Equation 1 or when the radiative rates are greater than collisional depopulation
out of the upper energy level of the transition.

There is also typically an increase in $n_{crit}^{thin, no\; bg}$ and $n_{eff}$ with
$J_u$ for a particular molecule (Figure 4).
This behavior can be understood from the functional dependences on $A_{jk}$
and $\gamma_{jk}$ in Equation 4
($n_{crit}^{thin, no\; bg}$ $\sim A_{jk}/\gamma_{jk}$). 
The Einstein $A$ for electric dipole allowed transitions is 
$A_{jk} \propto \mu_e^2 \nu_{jk}^3$, where $\mu_e$ is the molecular 
permanent electric dipole moment.  In general, molecules with larger $\mu_e$ will trace
denser gas. For instance, $\mu_e = 2.99$ D for HCN while $\mu_e = 0.11$ D for CO 
(Ebenstein \& Muenter 1984, Goorvitch 1994), partially explaining why HCN is a denser gas tracer 
than CO (the other effect being that the optical trapping in $^{12}$CO lines is often very large;
see \S2.2).
Since $\nu_{jk} \propto J_u$ and the Einstein $A_{jk}$ has a $\nu_{jk}^3$ dependence,
higher frequency transitions of a particular molecule 
will trace higher densities.  This statement is also true for comparing molecules with similar
permanent electric dipole moments but with transitions at different frequencies.
As an example, the frequency of the NH$_3$ ($\mu_e = 1.47$ D; Ueda \& Iwahori 1986) inversion transitions 
(J,K=J) are nearly a factor of $4$ lower than for the $1-0$ transitions of
\hcop\, \nthp\, and HCN, thus the spontaneous transition rate is more than $60$ times smaller,
lowering the critical densities of NH$_3$ (1,1) with respect to the 
$1-0$ transitions of \hcop\, \nthp\, and HCN.  One problem with the use of
$n_{crit}^{thin}$ becomes apparent for the NH$_3$ inversion transitions.  Since
NH$_3$ (1,1) through (4,4) are at nearly the same frequency ($23.7$ to $24.1$ GHz),
their $n_{crit}^{thin, no\; bg}$ are nearly identical.  $n_{eff}$ is sensitive to both line trapping
and $E_u/k$ of the transition resulting in an increase in $n_{eff}$ for NH$_3$ (1,1)
and (2,2) (N.B. \neff\ for NH$_3$ (3,3) and (4,4) are 
undefined at $T_k = 15$ K as we will discuss below).

Differences in collision rates also affect the ranking of molecules with density
in Figure 4.  HCN is observed to have 1 - 2 orders of magnitude 
higher $n_{crit}^{thin, no\; bg}$ and $n_{eff}$ than HCO$^+$.
Both \hcop\ and HCN transitions have similar $A_{jk}$ values, but \hcop\ is an
ion and the collisional cross section for collisions with H$_2$ is larger
than for neutral species (see Flower 1999 vs. Dumouchel et al. 2010).  
This reduces the critical density and effective 
excitation density of \hcop\ relative to HCN.

%%%%%%%%%%%%%%%%%% Figure 4  %%%%%%%%%%%%%%%%%%%%%%%%%

\begin{figure}
\figurenum{4}
\epsscale{1.0}
\plotone{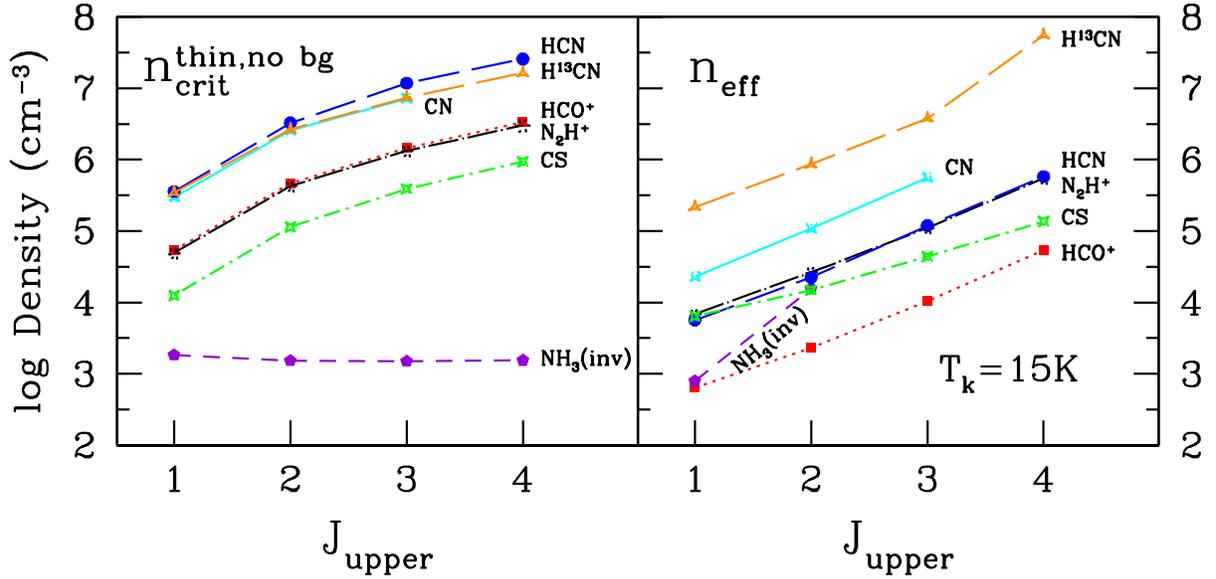}
\figcaption{ 
Left: $n_{crit}^{thin, no\; bg}$ calculated at $T_k = 15$ K for popular dense gas
tracers.  Each point corresponds to the upper state rotational quantum number J of
the transition. The ordering of labels correspond to the ordering of the curves in increasing
density for
each molecule.  Note that the curves for 
\hcop\ and N$_2$H$^+$ nearly overlap as well as for CN and H$^{13}$CN.
Right: $n_{eff}$ calculated for $T_k = 15$ K and at the reference
column density determined from surveys of Galactic sources (see Table 1).  
Note that the curves for HCN and N$_2$H$^+$ nearly overlap.
The critical density is typically one to 
two orders of magnitude above $n_{eff}$ for these tracers.}
\end{figure}

The ordering of species with increasing density is different
for $n_{crit}^{thin, no\; bg}$ versus $n_{eff}$ among many of the molecules plotted in Figure 4.
A problem with using $n_{crit}^{thin, no\; bg}$
can be seen when \hcop\ and \nthp\ are compared.  
\hcop\ and \nthp\ are both ions
with similar $A_{jk}$ values for transitions and similar collisional cross-sections 
and thus have similar $n_{crit}^{thin, no\; bg}$; however,
it is common knowledge to observers
of nearby, low-mass dense cores that \nthp\ is a denser gas tracer than \hcop .
Due to chemical effects, \nthp\ typically has lower abundance and therefore lower column 
density than \hcop\ resulting in a higher \neff\ (see Equation 13).

\neff\ also does a better job than $n_{crit}^{thin, no\; bg}$  
of characterizing the difference in density traced by isotopologues
versus the main species.   
The ratio of $A_{jk}$ for the $1 \rightarrow 0$ transition 
between H$^{13}$CN and HCN is $0.92$.  
This accounts for the slightly lower $n_{crit}^{thin, no\; bg}$ for H$^{13}$CN versus HCN; 
however, we know that the isotopologues
should trace higher density gas than the main species because it is more optically thin
and has a lower column density (see Equation 13).  
The average ratio of \neff\ for H$^{13}$CN/HCN $1 \rightarrow 0$ 
is $41$ ($T_k \in [10,100]$ K).  This ratio is 
not exactly equal to the assumed isotope ratio and column density ratio of $50$
because of differences in $A_{jk}$, collisional rates, and line trapping
between the isotopologue and main species.

These examples illustrates important 
advantages in using \neff\ rather than the critical density;
however, using $n_{eff}$ has a significant disadvantage 
in that \neff\ is undefined when an integrated intensity of 1 K km/s cannot be observed.  
This is the case for the NH$_3$ (3,3) ($E_u/k = 123.5$ K) and higher inversion lines at
low gas kinetic temperature.
When $E_u/k \gg T_k$, the upper energy state cannot attain a high enough population
to result in large enough radiative rates to produce a $1$ K km/s line.  
High $J_u$ transitions for species such as NH$_3$, HCN, \hcop , and \nthp\ 
will be undefined at low gas kinetic temperatures $(T_k \lesssim 20$ K).

Another problem with the standard definition of $n_{eff}$ is that the 
more complex the spectrum of a molecule (e.g. a heavy asymmetric top
molecule), the less likely a particular transition will attain enough level population
to produce a $1$ K km/s emission line.  For instance, no transitions of 
HNCO have intensities greater than $1$ K km/s for the observed
median column density of $\log \rm{med}\{N\} = 12.6$ (Gerner et al. 2014).
$n_{eff}$ for complex organic species (i.e. CH$_3$CHO, CH$_3$OCH$_3$, etc.) will be
undefined.  One solution is to reduce the $1$ K km/s criterion,   
Ultimately, the choice for this criterion is arbitrary, but
unless all researchers agree to use the same criterion, $n_{eff}$ could
not be compared among different papers and for different species.
The upper right panel of Figure 3 shows the non-linear sensitivity
of the integrated intensity for different transitions of \hcop\ 
with density.  As the integrated intensity criterion is lowered, the
curves flatten with $n_c$ indicating that a smaller decrease in the criterion results
in a larger decrease in $n_{eff}$.  
These problems limit the utility of $n_{eff}$ compared to $n_{crit}^{thin, no\; bg}$ which
can always be calculated if collision rates are known.

\section{A Caveat About Continuum Radiation Backgrounds}

We ignored continuum background radiation fields in \S2 when deriving the
expressions for $n_{crit}^{thin, no\; bg}$ and $n_{crit}^{thick, no\; bg}$,
while only the CMB was considered as a background in the RADEX models in \S3.
In this section we consider when contributions from the CMB and dust
continuum background are important.

The ratio of the rate of stimulated 
emission from $j \rightarrow k$ due to the CMB to spontaneous emission is given by
\begin{equation}
\frac{B_{jk} u_{\nu_{jk}}(T_{cmb})}{A_{jk}} = \frac{1}{e^{h\nu_{jk}/kT_{cmb}} - 1} = n_{ph}(T_{cmb},\nu_{jk}) \;\; .
\end{equation}
This ratio is $2$ for the NH$_3$ (J,K=J) inversion transitions and is a modest $0.26$  
at 90 GHz (3.3 mm), near the frequency of the fundamental transitions of commonly
observed dense gas tracers (e.g. \hcop , HCN, HNC, \nthp ).  This ratio
drops below $0.1$ for transitions with $\nu_{jk} > 140$ GHz ($\lambda_{jk} < 2.1$ mm).
Table 1 lists $n_{ph}(T_{cmb},\nu_{jk})$ for each transition.
%Draine (2011) advocates for a form of the optically thin critical density that includes
%a correction for stimulated emission from the CMB (ignoring the stimulated
%absorption term in Equation 1),
%\begin{equation}
%n_{crit}^{thin} = \frac{A_{jk} + B_{jk} u_{\nu_{jk}}(T_{cmb})}{\sum_{i \neq j} \gamma_{ji}} = 
%\frac{ A_{jk} [1 + n_{ph}(T_{cmb},\nu_{jk})] }{\sum_{i \neq j} \gamma_{ji}} \;\; .
%\end{equation}
%The CMB correction term in brackets in Equation 13 is given for each transition 
%in Table 1.
%If we include the stimulated emission term for the background 
%in Equation 4 and assume $\bar{\beta} \rightarrow 1$ (ignoring stimulated
%absorption), then we find that
%\begin{eqnarray}
%n_{crit}^{thin, cmb}  & = &  \frac{ A_{jk} + A_{jk} n_{ph}(T_{cmb},\nu_{jk})}
%{\sum_{i \neq j} \gamma_{ji} }
%\nonumber \\
% & = &
%\frac{ A_{jk} \left[ 1 + n_{ph}(T_{cmb},\nu_{jk})  \right] }
%{\sum_{i \neq j} \gamma_{ji}  }  \;\;.
%\end{eqnarray}
For low frequency (long wavelength) transitions, stimulated emission and absorption 
due to the CMB should not be ignored.

Interesting examples of the interaction of low frequency transitions with the 
CMB are the K-doublet transitions of ortho-H$_2$CO 
($1_{1,0} - 1_{1,1}$, $2_{1,1} - 2_{1,2}$, etc.).  The 
$4.6$ GHz ($6.5$ cm) $1_{1,0} - 1_{1,1}$ transition has a stimulated emission
rate to spontaneous emission rate ratio of $11.3$.  
Since the statistical weights of the two level are 
identical, the absorption rate from the lower level is identical
to the stimulated emission rate from the upper level.
This line is typically observed in absorption for 
densities $n \lesssim 10^{5}$ cm$^{-3}$ (Mangum et al. 2008).
The absorption is created by the lowest energy level ($1_{1,1}$) having 
a collisionally-induced overpopulation that results in absorption against the CMB 
($T_{ex} < T_{cmb}$)
over a wide range of physical conditions (Evans 1975, Garrison 1975).  
The calculation of $n_{crit}^{thin,no\; bg}$ gives a ridiculously low value of 
$13$ cm$^{-3}$ at $T_k = 10$ K.  
Clearly in cases such as these, the simple definition of critical density fails
and more sophisticated radiative transfer and statistical equilibrium calculations
that include the CMB background are required.

In addition to the CMB, dust continuum backgrounds from the Galactic interstellar
radiation field
and from emission local to the cloud can be important in modifying the level
populations.  The main effect of dust continuum emission on our definition
of critical density is to introduce a new stimulated rate out of level $j$
to higher energy levels, $i$. 
The energy density is now the energy density for optically thin dust emission
\begin{equation}
u_{\nu_{ij}}^{dust}(T_{dust}) = \tau_{\nu_{ij}}^{dust} u_{\nu_{ij}}(T_{dust}) \;\; ,
\end{equation}
where $\tau_{\nu_{ij}}^{dust}$ is the dust optical depth at
the frequency of the $i \rightarrow j$ transition and
$T_{dust}$ is the isothermal dust temperature (see Equation 4 of Shirley et al. 2011 for
a derivation of the equivalent isothermal dust temperature at a given frequency 
in sources with density and temperature gradients).
We only need to consider transitions 
between $i \rightarrow j$ that are allowed by electric dipole selection rules 
because the Einstein $B_{ji}$ for absorption of $j \rightarrow i$
is directly proportional to the Einstein $A_{ij}$ for spontaneous emission
of $i \rightarrow j$ (see Equation 2). 
For many simple molecules (\hcop , \nthp , HCN, CS, NH$_3$, etc.) 
there is only one electric dipole 
permitted transition from $j$ into the upper state $i$ that needs
to be considered.
The ratio of the rate at which this absorption from $j \rightarrow i$
occurs to the spontaneous rate
for the original $j \rightarrow k$ transition is
\begin{eqnarray}
\frac{B_{ji} u_{\nu_{ij}}^{dust}}{A_{jk}} & = & 
\frac{g_i}{g_j}\frac{A_{ij}}{A_{jk}}\frac{\tau_{\nu_{ij}}^{dust}}{\exp(h\nu_{ij}/kT_{dust}) - 1}
\nonumber \\
& = & 
\frac{g_i}{g_j}\frac{A_{ij}}{A_{jk}} \mu m_{\rm{H}} N_{\rm{H}_2} \kappa_{\nu_{jk}} 
n_{ph}(T_{dust}, \nu_{ij}) \;\; ,
\end{eqnarray}
where $\mu = 2.8$ is the mean mass per H$_2$ 
molecule (see Kauffmann et al. 2008), $N_{\rm{H}_2}$ is the H$_2$ column density (cm$^{-2}$),
and $\kappa_{\nu_{jk}}$ is dust opacity (cm$^2$/gram of gas) at the frequency
of the $j \rightarrow i$ absorption.  
The ratio of rates in Equation 16 is linear proportional
to the H$_2$ column density and increases non-linearly with dust temperature.

For the following examples, we calculate the ratio of rates in Equation 16
assuming a H$_2$ column density of $10^{22}$ cm$^{-3}$ (corresponding
to $A_V \sim 10$ mag), a gas mass to dust mass
ratio of $100:1$, and Ossenkopf \& Henning opacities for coagulated grains
with thin ice mantles (OH5, Ossenkopf \& Henning 1994) that are
appropriate for dense cores.
For the NH$_3$ (1,1) transition, the upper level is radiatively coupled to
the $J = 2, K = 1$ level via a 1.168 THz ($267$ $\mu$m) far-infrared transition (see Figure 1).
If NH$_3$ is bathed in a dust continuum of $10$ K, 
the stimulated rate of absorption out of the
upper inversion level is four times the spontaneous rate of the (1,1) transition.
%The ratio increases to $500$ if the dust temperature is $50$ K.
This high ratio is partially driven by the ratio of
spontaneous rates between the terahertz transition and the centimeter inversion transition 
which is $A_{ij}/A_{jk} = 7.2 \times 10^5$.   
The net effect is that \neff\ becomes larger as the energy density from dust continuum 
increases. 
When the upper level of a low frequency transition is radiatively coupled to far-infrared
transitions, the absorption from dust continuum is an important
rate that should not be ignored.  
This is the case for the centimeter transitions of molecules such as NH$_3$, OH, and CH.

The effects of dust continuum radiation are significantly less 
for the lowest energy transitions of simple linear molecules.
For the $267.5$ GHz (1.1 mm) \hcop\ $3-2$ transition at $10$ K, the
ratio of rates is $6 \times 10^{-4}$.  
At 50 K, the ratio of rates rises to a meager $7 \times 10^{-3}$.  
These very low ratios indicate
that the absorption due to dust continuum emission
is not important for the level populations associated with 
millimeter transitions of linear molecules such as HCO$^+$, HCN, CS, and \nthp .

We have only considered the rotational
levels of the molecule, but there can also be coupling between the radiation
field and vibrational levels or electronic levels 
of the molecule that can effect the level populations
of the ground vibrational state rotational levels.  A mid-infrared
continuum can pump the molecule from the ground vibrational state into
an excited vibrational state which then radiatively decays via ro-vibrational
transitions.  For example, the HCN and HNC molecules have bending modes 
($\nu_1,\nu_2,\nu_3 = 0,1,0$) that can be excited
by mid-infrared pumping at $14$ $\mu$m and $21$ $\mu$m respectively.  
In environments where such pumping may be important (i.e. the inner envelope of
high-mass protostars or near an AGN), the level populations can be affected
by mid-infrared pumping.  In these cases, more sophisticated radiative
transfer models are needed.

\section{Summary}

In this tutorial, $n_{crit}^{thin, no\; bg}$ and \neff\ are defined and their
properties analyzed.  The values for $n_{crit}^{thin, no\; bg}$ and \neff\ are
calculated for multiple transitions of 
12 commonly observed dense molecular gas species in Table 1.
%The relative ordering in density of some species and transitions can be understood
%from the function dependence of $n_{crit}^{thin, no\; bg} \sim A_{jk}/\gamma_{jk}$
%with $A_{jk} \propto \mu_e^2 \nu_{jk}^3$ for electric dipole transitions. 
\neff\ inversely depends on the choice of column density and increases non-linearly
as gas kinetic temperatures are lowered ($T_k < 20$ K).  We have estimated
the reference column density for difference species listed in Table 1 from surveys of 
cores and clumps
in the Milky Way.  The \neff\ quoted can be easily scaled to another column density
using Equation 13.

Since the definition of \neff\ is arbitrary (the $1$ K km/s criteria) and does not
depend on a specific ratio of radiative rates to collisional depopulation rates
(c.f. $\Re = 1$ for the definition of $n_{crit}$), 
it is seductive to use $n_{crit}^{thin, no\; bg}$ to characterize
the density at which a transition is excited as it
can be analytically calculated; however, $n_{crit}^{thin, no\; bg}$ does
not account for radiative trapping which can easily lower the effective critical 
density by more than an order of magnitude for commonly observed dense gas tracers
such as \hcop\ $1-0$, HCN $1-0$, \nthp\ $1-0$, etc.
The combination of $\Re > 1$ and radiative trapping that generate a $1$ K km/s line
lower \neff\ by $\sim 1 - 2$
orders of magnitude relative to $n_{crit}^{thin, no\; bg}$ for many commonly
observed millimeter and submillimeter lines.
Furthermore, \neff\ can account for abundance or column density differences
to better differentiate the excitation density of different species than
$n_{crit}^{thin, no\; bg}$ (i.e. \hcop\ vs. \nthp\ or isotpologues vs. the main species).
Overall, the effective excitation density gives a better estimate of the 
range of densities at which a modest (1 K km/s) molecular line can be observed; 
however, use of $n_{eff}$ has the disadvantages that an appropriate column density must be
determined for the observations of interest and that it is not defined for all transitions 
(in particular when $E_u/k \gg T_k$) and for all species (e.g. HNCO, CH$_3$CHO, etc.).

$n_{crit}^{thin, no\; bg}$ and \neff\ provide only rough
estimates of the densities traced and should not be over-interpreted.
More sophisticated tools such as the Contribution Function
(Tafalla et al. 2006) may be used to determine the various contributions to
the observed line profile along the line of sight (see Pavlyuchenkov et al. 2008
for a detailed analysis of these techniques).
Ultimately, if one wants to understand their observed spectra, radiative transfer
modeling with publicly available codes are a fast and efficient way to determine the 
physical properties of a region.  
%Appendix C of the RADEX summary paper displays
%several diagnostic plots derived from models of line ratios for many
%commonly observed molecular dense gas tracers (van der Tak et al. 2007).

\section*{Acknowledgments}

I am grateful to the Max-Planck-Institut f\"ur Astronomie and Thomas Henning 
for hosting my Sabbatical during which this note was written.  
I would also like to thank John Black, Lee Mundy, Neal Evans, Henrik Beuther, 
Cecilia Ceccarelli, J.D. Smith, and the anonymous referee for 
comments and useful discussions that improved this manuscript.

%%%%%%%%%%%%%%%%%%%%%% References %%%%%%%%%%%%%%%%%%%%%%%%%%%%%%%%%%

%comment next line to include figures
%\end{document}

%%%%%%%%%%%%%%%%%% Figures %%%%%%%%%%%%%%%%%%%%%%%%%%%%%%%%%%

%%%%%%%%%%%%%%%%%% Tables %%%%%%%%%%%%%%%%%%%%%%%%%%%%%%%%%%

%%%%%%%%%%%%%%%%%% Table 1 - Summary of $n_{crit}^{thin, no\; bg}$ and $n_{eff}$ for Dense Gas Tracers%%%%%%%
\begin{deluxetable}{llrrclccccccccccr}
\rotate
\tabletypesize{\tiny}
\tablecolumns{17}
\tablecaption{$n_{crit}^{thin}$ and $n_{eff}$ for Dense Gas Tracers with Collisions with H$_2$ \label{tab1}}
\tablewidth{0pt}
\tablehead{
\colhead{Molecule} &
\colhead{$j \rightarrow k$} &
\colhead{$\nu_{jk}$} &
\colhead{$E_j/\rm{k}$} &
\colhead{$A_{jk}$} &
\colhead{$n_{ph}(T_{cmb})$} &
\multicolumn{4}{c}{\underline{$\;\;\;\;\;\;\;\;\;\;\;\;$ $n_{crit}^{thin, no\; bg}(T_k)$ $\;$ \rm{cm}$^{-3}$ $\;\;\;\;\;\;\;\;\;\;\;\;$}} &
\colhead{$\log N_{ref}$} &
\multicolumn{5}{c}{\underline{$\;\;\;\;\;\;\;\;\;\;\;\;\;\;\;\;\;\;\;\;\;$ $n_{eff}(T_k,N_{ref})$\tablenotemark{a} $\;$ \rm{cm}$^{-3}$ $\;\;\;\;\;\;\;\;\;\;\;\;\;\;\;\;$}} &
\colhead{$\gamma_{ji}$\tablenotemark{b}} \\
\colhead{} &
\colhead{} &
\colhead{(GHz)} &
\colhead{(K)} &
\colhead{(s$^{-1}$)} &
\colhead{} &
\colhead{10K} &
\colhead{20K} &
\colhead{50K} &
\colhead{100K} &
\colhead{(cm$^{-2}$)} &
\colhead{10K} &
\colhead{15K} &
\colhead{20K} &
\colhead{50K} &
\colhead{100K} &
\colhead{Ref.}
}
\startdata
HCO$^+$	&	$1-0$	&	89.189	&	4.28	&	4.3E-5	&	0.264	&	6.8E+4	&	4.5E+4	&	2.9E+4	&	2.3E+4	& 	14.0	&	9.5E+2	&	6.4E+2	&	5.3E+2	& 	3.3E+2	&	2.6E+2	&	1	\\
	&	$2-1$	&	178.375	&	12.84	&	4.1E-4	&	0.046	&	5.6E+5	&	4.2E+5	&	2.8E+5	&	2.2E+5	& 		&	4.1E+3	&	2.3E+3	&	1.7E+3	& 	8.9E+2	&	6.5E+2	&		\\
	&	$3-2$	&	267.558	&	25.68	&	1.5E-3	&	0.009	&	1.6E+6	&	1.4E+6	&	1.0E+6	&	8.1E+5	& 		&	2.4E+4	&	1.1E+4	&	6.8E+3	& 	2.6E+3	&	1.7E+3	&		\\
	&	$4-3$	&	356.734	&	42.80	&	3.6E-3	&	0.002	&	3.6E+6	&	3.2E+6	&	2.5E+6	&	2.0E+6	& 		&	1.9E+5	&	5.4E+4	&	2.8E+4	& 	7.2E+3	&	4.0E+3	&		\\
H$^{13}$CO$^+$	&	$1-0$	&	86.754	&	4.16	&	3.9E-5	&	0.279	&	6.2E+4	&	4.1E+4	&	2.7E+4	&	2.0E+4	& 	12.3	&	3.9E+4	&	2.7E+4	&	2.2E+4	& 	1.4E+4	&	1.1E+4	&	1	\\
	&	$2-1$	&	173.507	&	12.49	&	3.7E-4	&	0.050	&	5.1E+5	&	3.8E+5	&	2.6E+0	&	2.0E+5	& 		&	1.7E+5	&	9.6E+4	&	7.1E+4	& 	3.7E+4	&	2.6E+4	&		\\
	&	$3-2$	&	260.255	&	24.98	&	1.3E-3	&	0.011	&	1.5E+6	&	1.3E+6	&	9.5E+5	&	7.3E+5	& 		&	1.3E+6	&	4.3E+5	&	2.7E+5	& 	9.6E+4	&	6.2E+4	&		\\
	&	$4-3$	&	346.998	&	41.63	&	3.3E-3	&	0.002	&	3.4E+6	&	2.9E+6	&	2.3E+6	&	1.8E+6	& 		&	\nodata	&	2.7E+6	&	1.1E+6	& 	2.7E+5	&	1.6E+5	&		\\
N$_2$H$^+$	&	$1-0$	&	93.174	&	4.47	&	3.6E-5	&	0.242	&	6.1E+4	&	4.1E+4	&	2.6E+4	&	2.0E+4	& 	13.0	&	1.0E+4	&	6.7E+3	&	5.5E+3	& 	3.4E+3	&	2.6E+3	&	1,2	\\
	&	$2-1$	&	186.345	&	13.41	&	3.5E-4	&	0.040	&	5.0E+5	&	3.7E+5	&	2.6E+5	&	1.9E+5	& 		&	4.6E+4	&	2.6E+4	&	1.9E+4	& 	9.7E+3	&	6.9E+3	&		\\
	&	$3-2$	&	279.512	&	26.83	&	1.3E-3	&	0.007	&	1.4E+6	&	1.2E+6	&	9.2E+5	&	7.1E+5	& 		&	2.6E+5	&	1.1E+5	&	6.8E+4	& 	2.5E+4	&	1.6E+4	&		\\
	&	$4-3$	&	372.673	&	44.71	&	3.1E-3	&	0.001	&	3.2E+6	&	2.8E+6	&	2.2E+6	&	1.7E+6	& 		&	3.0E+6	&	5.4E+5	&	2.7E+5	& 	6.3E+4	&	3.4E+4	&		\\
HCN	&	$1-0$	&	88.632	&	4.25	&	2.4E-5	&	0.268	&	4.7E+5	&	3.0E+5	&	1.7E+5	&	1.1E+5	& 	14.0	&	8.4E+3	&	5.6E+3	&	4.5E+3	& 	2.6E+3	&	1.7E+3	&	3	\\
	&	$2-1$	&	177.261	&	12.76	&	2.3E-4	&	0.047	&	4.1E+6	&	2.8E+6	&	1.6E+6	&	1.1E+6	& 		&	4.1E+4	&	2.2E+4	&	1.6E+4	& 	7.3E+3	&	4.5E+3	&		\\
	&	$3-2$	&	265.886	&	25.52	&	8.4E-4	&	0.010	&	1.4E+7	&	1.0E+7	&	5.7E+6	&	3.8E+6	& 		&	3.0E+5	&	1.2E+5	&	7.3E+4	& 	2.5E+4	&	1.3E+4	&		\\
	&	$4-3$	&	354.505	&	42.53	&	2.1E-3	&	0.002	&	3.0E+7	&	2.3E+7	&	1.4E+7	&	9.1E+6	& 		&	2.1E+6	&	5.8E+5	&	3.2E+5	& 	8.1E+4	&	3.7E+4	&		\\
H$^{13}$CN	&	$1-0$	&	86.340	&	4.14	&	2.2E-5	&	0.282	&	5.3E+5	&	2.5E+5	&	1.3E+5	&	9.7E+4	& 	12.3	&	3.5E+5	&	2.2E+5	&	1.6E+5	& 	8.7E+4	&	6.5E+4	&	4	\\
	&	$2-1$	&	172.678	&	12.43	&	2.1E-4	&	0.051	&	3.4E+6	&	2.2E+6	&	1.2E+6	&	9.1E+5	& 		&	1.9E+6	&	8.6E+5	&	5.8E+5	& 	1.8E+5	&	1.7E+5	&		\\
	&	$3-2$	&	259.012	&	24.86	&	7.7E-4	&	0.011	&	8.8E+6	&	6.6E+6	&	4.1E+6	&	3.3E+6	& 		&	2.5E+7	&	3.8E+6	&	2.0E+6	& 	6.3E+5	&	4.2E+5	&		\\
	&	$4-3$	&	345.340	&	41.43	&	1.9E-3	&	0.002	&	1.9E+7	&	1.5E+7	&	9.7E+6	&	7.7E+6	& 		&	\nodata	&	5.6E+7	&	1.1E+7	& 	1.9E+6	&	1.1E+6	&		\\
HNC	&	$1-0$	&	90.664	&	4.35	&	2.7E-5	&	0.256	&	1.4E+5	&	1.1E+5	&	8.4E+4	&	7.0E+4	& 	14.0	&	3.7E+3	&	2.7E+3	&	2.3E+3	& 	1.7E+3	&	1.3E+3	&	3	\\
	&	$2-1$	&	181.325	&	13.05	&	2.6E-4	&	0.043	&	1.3E+6	&	1.0E+6	&	8.2E+5	&	6.8E+5	& 		&	2.0E+4	&	1.2E+4	&	9.3E+3	& 	5.5E+3	&	3.9E+3	&		\\
	&	$3-2$	&	271.981	&	26.11	&	9.3E-4	&	0.009	&	5.1E+6	&	4.0E+6	&	3.1E+6	&	2.5E+6	& 		&	1.6E+5	&	7.4E+4	&	4.9E+4	& 	2.1E+4	&	1.2E+4	&		\\
	&	$4-3$	&	362.630	&	43.51	&	2.3E-3	&	0.002	&	1.3E+7	&	1.0E+7	&	7.8E+6	&	6.2E+6	& 		&	1.4E+6	&	4.2E+5	&	2.4E+5	& 	6.8E+4	&	3.4E+4	&		\\
CN	&	$1_{3/2}-0_{1/2}$	&	113.495	&	5.45	&	1.2E-5	&	0.158	&	4.1E+5	&	2.4E+5	&	1.1E+5	&	6.4E+4	& 	14.0	&	3.8E+4	&	2.3E+4	&	1.7E+4	& 	7.7E+3	&	4.6E+3	&	5	\\
	&	$2_{5/2}-1_{3/2}$	&	226.876	&	16.34	&	1.1E-4	&	0.019	&	3.2E+6	&	2.2E+6	&	1.1E+6	&	6.4E+5	& 		&	2.3E+5	&	1.1E+5	&	7.3E+4	& 	2.8E+4	&	1.5E+4	&		\\
	&	$3_{7/2}-2_{5/2}$	&	340.249	&	32.66	&	4.1E-4	&	0.003	&	8.3E+6	&	6.5E+6	&	3.5E+6	&	2.1E+6	& 		&	1.8E+6	&	5.6E+5	&	3.6E+5	& 	1.1E+5	&	5.1E+4	&		\\
CS	&	$1-0$	&	48.991	&	2.40	&	1.7E-6	&	0.734	&	1.5E+4	&	1.1E+4	&	7.8E+3	&	5.7E+3	& 	13.5	&	8.7E+3	&	6.4E+3	&	5.4E+3	& 	3.4E+3	&	2.4E+3	&	6	\\
	&	$2-1$	&	97.981	&	7.10	&	1.7E-5	&	0.218	&	1.3E+5	&	1.0E+5	&	7.4E+4	&	5.5E+4	& 		&	2.2E+4	&	1.5E+4	&	1.2E+4	& 	6.9E+3	&	4.7E+3	&		\\
	&	$3-2$	&	146.969	&	14.10	&	6.1E-5	&	0.082	&	4.4E+5	&	3.6E+5	&	2.6E+5	&	2.0E+5	& 		&	7.4E+4	&	4.4E+4	&	3.3E+4	& 	1.6E+4	&	1.0E+4	&		\\
	&	$4-3$	&	195.954	&	23.50	&	1.5E-4	&	0.033	&	1.0E+6	&	8.6E+5	&	6.4E+5	&	4.8E+5	& 		&	2.8E+5	&	1.4E+5	&	9.2E+4	& 	3.6E+4	&	2.0E+4	&		\\
	&	$5-4$	&	244.936	&	35.30	&	3.0E-4	&	0.014	&	2.0E+6	&	1.7E+6	&	1.3E+6	&	9.5E+5	& 		&	1.4E+6	&	4.2E+5	&	2.5E+5	& 	7.6E+4	&	3.8E+4	&		\\
	&	$6-5$	&	293.912	&	49.40	&	5.2E-4	&	0.006	&	3.4E+6	&	2.9E+6	&	2.2E+6	&	1.6E+6	& 		&	\nodata	&	1.5E+6	&	7.1E+5	& 	1.6E+5	&	7.2E+4	&		\\
	&	$7-6$	&	342.883	&	65.80	&	8.4E-4	&	0.002	&	5.0E+6	&	4.4E+6	&	3.4E+6	&	2.6E+6	& 		&	\nodata	&	8.8E+6	&	2.1E+6	& 	3.4E+5	&	1.3E+5	&		\\
p-NH$_3$\tablenotemark{c}	&	$(1,1)$	&	23.694	&	1.14	&	1.7E-7	&	1.940	&	2.0E+3	&	1.8E+3	&	1.2E+3	&	8.7E+2	& 	14.3	&	9.3E+2	&	7.9E+2	&	7.6E+2	& 	7.4E+2	&	7.3E+2	&	7	\\
	&	$(2,2)$	&	23.723	&	42.32	&	2.3E-7	&	1.937	&	1.6E+3	&	1.5E+3	&	1.4E+3	&	1.1E+3	& 		&	\nodata	&	1.6E+4	&	3.9E+3	& 	1.3E+3	&	9.4E+2	&		\\
	&	$2_1^+-1_1^-$	&	1168.452	&	57.21	&	1.2E-2	&	0.000	&	1.2E+8	&	1.1E+8	&	9.0E+7	&	6.7E+7	& 		&	\nodata	&	4.8E+7	&	8.6E+6	& 	1.4E+6	&	8.2E+5	&		\\
	&	$2_1^--1_1^+$	&	1215.245	&	58.32	&	1.4E-2	&	0.000	&	1.4E+8	&	1.2E+8	&	1.0E+8	&	7.5E+7	& 		&	\nodata	&	7.2E+7	&	1.0E+7	& 	1.6E+6	&	8.9E+5	&		\\
o-NH$_3$\tablenotemark{c}	&	$(3,3)$	&	23.870	&	123.54	&	2.6E-7	&	1.923	&	1.5E+3	&	1.5E+3	&	1.4E+3	&	1.2E+3	& 	14.6	&	\nodata	&	\nodata	&	\nodata	& 	4.0E+2	&	1.7E+2	&	7	\\
	&	$1_0-0_0$	&	572.498	&	27.48	&	1.6E-3	&	0.000	&	4.0E+7	&	3.1E+7	&	1.7E+7	&	9.9E+6	& 		&	6.2E+5	&	1.5E+5	&	8.4E+4	& 	3.1E+4	&	2.5E+4	&		\\
	&	$2_0-1_0$	&	1214.859	&	85.78	&	1.8E-2	&	0.000	&	2.1E+8	&	1.9E+8	&	1.6E+8	&	1.1E+8	& 		&	\nodata	&	4.2E+7	&	4.4E+6	& 	5.2E+5	&	3.0E+5	&		\\
p-H$_2$CO	&	$1_{0,1}-0_{0,0}$	&	72.838	&	3.50	&	8.2E-6	&	0.386	&	4.5E+4	&	2.8E+4	&	1.8E+4	&	1.4E+4	& 	12.7	&	5.0E+4	&	3.2E+4	&	2.6E+4	& 	1.9E+4	&	1.8E+4	&	8	\\
	&	$2_{0,2}-1_{0,1}$	&	145.603	&	10.50	&	7.8E-5	&	0.084	&	3.4E+5	&	2.5E+5	&	1.7E+5	&	1.3E+5	& 		&	1.5E+5	&	8.2E+4	&	6.3E+4	& 	4.3E+4	&	4.0E+4	&		\\
	&	$3_{0,3}-2_{0,2}$	&	218.222	&	21.00	&	2.8E-4	&	0.022	&	9.7E+5	&	7.8E+5	&	5.7E+5	&	4.7E+5	& 		&	7.7E+5	&	2.9E+5	&	2.0E+5	& 	1.2E+5	&	1.1E+5	&		\\
	&	$4_{0,4}-3_{0,3}$	&	290.623	&	34.90	&	6.9E-4	&	0.006	&	1.9E+6	&	1.7E+6	&	1.3E+6	&	1.1E+6	& 		&	\nodata	&	1.5E+6	&	7.3E+5	& 	3.1E+5	&	2.6E+5	&		\\
	&	$5_{0,5}-4_{0,4}$	&	362.736	&	52.30	&	1.4E-3	&	0.002	&	3.6E+6	&	3.1E+6	&	2.6E+6	&	2.2E+6	& 		&	\nodata	&	\nodata	&	3.8E+6	& 	8.0E+5	&	5.9E+5	&		\\
o-H$_2$CO	&	$1_{1,0}-1_{1,1}$	&	4.830	&	15.40	&	3.6E-9	&	11.310	&	1.3E+1	&	1.0E+1	&	7.6E+0	&	6.1E+0	& 	13.0	&	\nodata	&	\nodata	&	\nodata	& 	\nodata	&	\nodata	&	8	\\
	&	$2_{1,1}-2_{1,2}$	&	14.488	&	22.60	&	3.2E-8	&	3.456	&	1.1E+2	&	8.7E+1	&	6.5E+1	&	5.3E+1	& 		&	\nodata	&	\nodata	&	\nodata	& 	\nodata	&	\nodata	&		\\
	&	$2_{1,2}-1_{1,1}$	&	140.840	&	21.90	&	5.3E-5	&	0.092	&	1.5E+5	&	1.3E+5	&	1.1E+5	&	9.0E+4	& 		&	9.5E+4	&	5.9E+4	&	4.6E+4	& 	2.6E+4	&	2.0E+4	&		\\
	&	$2_{1,1}-1_{1,0}$	&	150.498	&	22.60	&	6.5E-5	&	0.077	&	2.2E+5	&	1.8E+5	&	1.3E+5	&	1.1E+5	& 		&	1.7E+5	&	1.1E+5	&	8.6E+4	& 	4.9E+4	&	3.5E+4	&		\\
	&	$3_{1,3}-2_{1,2}$	&	211.211	&	32.10	&	2.3E-4	&	0.025	&	5.6E+5	&	5.1E+5	&	4.3E+5	&	3.8E+5	& 		&	4.6E+5	&	2.0E+5	&	1.4E+5	& 	6.3E+4	&	4.9E+4	&		\\
	&	$3_{1,2}-2_{1,1}$	&	225.698	&	33.40	&	2.8E-4	&	0.019	&	8.4E+5	&	7.0E+5	&	5.5E+5	&	4.6E+5	& 		&	8.9E+5	&	3.8E+5	&	2.6E+5	& 	1.3E+5	&	9.0E+4	&		\\
	&	$4_{1,4}-3_{1,3}$	&	281.527	&	45.60	&	5.9E-4	&	0.007	&	1.4E+6	&	1.3E+6	&	1.1E+6	&	9.8E+5	& 		&	2.9E+7	&	9.9E+5	&	5.1E+5	& 	1.7E+5	&	1.2E+5	&		\\
	&	$4_{1,3}-3_{1,2}$	&	300.837	&	47.90	&	7.2E-4	&	0.005	&	2.0E+6	&	1.7E+6	&	1.4E+6	&	1.2E+6	& 		&	\nodata	&	1.9E+6	&	9.6E+5	& 	3.3E+5	&	2.2E+5	&		\\
	&	$5_{1,5}-4_{1,4}$	&	351.769	&	62.50	&	1.2E-3	&	0.002	&	2.8E+6	&	2.6E+6	&	2.2E+6	&	2.0E+6	& 		&	\nodata	&	1.7E+7	&	2.3E+6	& 	4.3E+5	&	2.6E+5	&		\\
HC$_3$N\tablenotemark{d}	&	$2-1$	&	18.196	&	1.31	&	3.9E-7	&	2.659	&	9.7E+2	&	7.5E+2	&	5.3E+2	&	4.1E+2	& 	13.0	&	5.9E+3	&	3.1E+3	&	2.3E+3	& 	1.0E+3	&	8.4E+2	&	9	\\
	&	$3-2$	&	27.294	&	2.62	&	1.4E-6	&	1.628	&	3.5E+3	&	2.6E+3	&	1.8E+3	&	1.6E+3	& 		&	4.7E+3	&	3.0E+3	&	2.3E+3	& 	1.1E+3	&	8.7E+2	&		\\
	&	$4-3$	&	36.392	&	4.37	&	3.5E-6	&	1.119	&	9.1E+3	&	6.8E+3	&	4.7E+3	&	3.7E+3	& 		&	6.2E+3	&	3.8E+3	&	2.8E+3	& 	1.5E+3	&	1.1E+3	&		\\
	&	$5-4$	&	45.490	&	6.55	&	6.9E-6	&	0.819	&	1.8E+4	&	1.3E+4	&	9.6E+3	&	7.2E+3	& 		&	9.6E+3	&	5.6E+3	&	4.1E+3	& 	2.1E+3	&	1.5E+3	&		\\
	&	$8-7$	&	72.784	&	15.72	&	2.9E-5	&	0.387	&	7.7E+4	&	5.9E+4	&	4.1E+4	&	3.1E+4	& 		&	5.3E+4	&	2.5E+4	&	1.6E+4	& 	6.4E+3	&	4.0E+3	&		\\
	&	$9-8$	&	81.881	&	19.65	&	4.2E-5	&	0.312	&	1.1E+5	&	8.3E+4	&	6.2E+4	&	4.6E+4	& 		&	1.1E+5	&	4.2E+4	&	2.6E+4	& 	9.4E+3	&	5.7E+3	&		\\
	&	$10-9$	&	90.979	&	24.01	&	5.8E-5	&	0.254	&	1.6E+5	&	1.2E+5	&	8.2E+4	&	6.3E+4	& 		&	4.3E+5	&	7.2E+4	&	4.3E+4	& 	1.4E+4	&	8.1E+3	&		\\
	&	$11-10$	&	100.076	&	28.82	&	7.8E-5	&	0.209	&	2.0E+5	&	1.6E+5	&	1.1E+5	&	9.2E+4	& 		&	\nodata	&	1.6E+5	&	7.0E+4	& 	2.0E+4	&	1.1E+4	&		\\
	&	$12-11$	&	109.173	&	34.06	&	1.0E-4	&	0.173	&	2.8E+5	&	2.1E+5	&	1.5E+5	&	1.1E+5	& 		&	\nodata	&	2.3E+5	&	1.1E+5	& 	2.8E+4	&	1.6E+4	&		\\
	&	$14-13$	&	127.367	&	45.85	&	1.6E-4	&	0.120	&	3.4E+5	&	3.1E+5	&	2.5E+5	&	1.9E+5	& 		&	\nodata	&	\nodata	&	3.8E+5	& 	5.7E+4	&	3.0E+4	&		\\
	&	$15-14$	&	136.464	&	52.40	&	2.0E-4	&	0.100	&	5.5E+5	&	4.3E+5	&	3.1E+5	&	2.4E+5	& 		&	\nodata	&	\nodata	&	8.2E+5	& 	8.1E+4	&	4.2E+4	&		\\
CH$_3$CN\tablenotemark{e}	& 	$4_0-3_0$	&	73.590	&	8.83	&	3.2E-5	&	0.379	&	1.2E+5	&	8.6E+4	&	5.4E+4	&	4.1E+4	& 	13.0	&	5.2E+4	&	4.0E+4	&	3.7E+4	& 	2.5E+4	&	2.5E+4	&	10	\\
	& 	$5_0-4_0$	&	91.987	&	13.24	&	6.3E-5	&	0.249	&	2.5E+5	&	1.7E+5	&	1.1E+5	&	8.3E+4	& 		&	1.2E+5	&	8.4E+4	&	7.4E+4	& 	4.3E+4	&	4.0E+4	&		\\
	& 	$6_0-5_0$	&	110.384	&	18.54	&	1.1E-4	&	0.169	&	4.5E+5	&	3.1E+5	&	1.9E+5	&	1.5E+5	& 		&	3.2E+5	&	1.8E+5	&	1.5E+5	& 	7.6E+4	&	6.5E+4	&		\\
	& 	$7_0-6_0$	&	128.779	&	24.72	&	1.8E-4	&	0.117	&	7.3E+5	&	5.1E+5	&	3.1E+5	&	2.4E+5	& 		&	1.0E+6	&	4.2E+5	&	3.1E+5	& 	1.4E+5	&	1.1E+5	&		\\
	& 	$8_0-7_0$	&	147.175	&	31.79	&	2.7E-4	&	0.082	&	1.1E+6	&	7.7E+5	&	4.8E+5	&	3.6E+5	& 		&	8.8E+6	&	1.0E+6	&	6.5E+5	& 	2.3E+5	&	1.7E+5	&		\\
	& 	$9_0-8_0$	&	165.569	&	39.73	&	3.8E-4	&	0.058	&	1.6E+6	&	1.1E+6	&	6.9E+5	&	5.2E+5	& 		&	\nodata	&	3.1E+6	&	1.4E+6	& 	4.0E+5	&	2.8E+5	&		\\
\enddata
\tablenotetext{a}{ The effective excitation density may be converted to another column density using $n_{eff}(T_k,N) = n_{eff}(T_k,N_{ref}) \, \frac{N_{ref}}{N}$ for 
$n_{eff} \lesssim 5 \times 10^6$ cm$^{-3}$.  Logarithmic interpolation should be
used to convert $n_{eff}$ to other $T_k$.}
\tablenotetext{b}{ The collisional rates for collisions with H$_2$ 
in cm$^{-3}$ s$^{-1}$ were obtained from the Leiden LAMDA database (August 2014).  The original references for the collision rates are: 1. Flower et al. 1999, 2. Daniel et al. 2005, 3. Dumouchel et al. 2010, 4. Green \& Thaddeus 1974, 5. Lique et al. 2011, 6. Lique et al. 2006, 7. Danby et al. 1988, 8. Wiesenfeld \& Faure 2013, 9. Green \& Chapman 1978, 10. Green 1986.  Rates calculated for collisions with He are scaled in the LAMBDA database by the reduced mass (1.36) to convert to approximate rates for collisions with H$_2$.  Collision rates with hyperfine transitions collapsed were used for the calculations in this table.}
\tablenotetext{c}{ $n_{crit}^{thin, no\; bg}$ was logarithmically extrapolated to $T_k = 10$K since
the lowest $T_k$ calculated in the collision rate file was $15$K.}
\tablenotetext{d}{ $n_{crit}^{thin, no\; bg}$ was logarithmically extrapolated to $T_k = 100$K since
the highest $T_k$ calculated in the collision rate file was $80$K.}
\tablenotetext{e}{ $n_{crit}^{thin, no\; bg}$ was logarithmically extrapolated to $T_k = 10$K since the lowest $T_k$ calculated in the collision rate file was $20$K.}
\end{deluxetable}

\end{document}